\documentclass[aps,preprint,groupedaddress,showpacs,longbibliography,nofootinbib]{revtex4-2}
\usepackage{amssymb}
\usepackage[breaklinks,colorlinks,citecolor=green,urlcolor=blue,linkcolor=red]{hyperref}
\usepackage{graphicx}
\usepackage{epstopdf}
\usepackage{diagbox}
\usepackage{float}
\usepackage{subfigure}
\usepackage{multirow}
\usepackage{CJK}
\usepackage{tikz}
\usepackage{enumerate}

\begin{document}

\newcommand{\RN}[1]{\uppercase\expandafter{\romannumeral#1}}
\newcommand{\CPV}{$C\!P$V}
\newcommand{\CPA}{$C\!P$A}


\title{\boldmath Decay-angular-distribution correlated $C\!P$ violation in heavy hadron cascade decays}


\author{Yu-Jie Zhao}
\author{Zhen-Hua Zhang}
\affiliation{School of Nuclear Science and Technology, University of South China, Hengyang, 421001, Hunan, China}

\author{Xin-Heng Guo}
\affiliation{College of Nuclear Science and Technology, Beijing Normal University, Beijing 100875, China}


\date{\today}

\begin{abstract}
$C\!P$ violation in baryon decay processes is still undiscovered to date.
We present a general analysis of the decay-angular-distributions and the corresponding $C\!P$ asymmetries in cascade decays of the type $\mathbb{H}\to R(\to ab) c$, where $\mathbb{H}$ is a heavy hadron that decays through weak interactions $\mathbb{H}\to R c$, and the resonance $R$ decays strongly via $R\to ab$.
Based on the analysis, we propose to search for $C\!P$ violation in the decay-angular-distributions in the cascade decay processes ${\mathbb{B}}\to \mathcal{B} M$,  with $\mathcal{B}$ or $M$ subsequently decaying through strong interactions, where ${\mathbb{B}}$ is the mother baryon, $\mathcal{B}$ or $M$ are the daughter baryon and meson, respectively, and $M$ has to be spin-nonzero.
We also present some typical decay channels in which the search for such kinds of $C\!P$ asymmetries can be performed.
\end{abstract}
\maketitle
\section{introduction}
In the Observable Universe, the density of baryons is by far larger than that of anti-baryons \cite{Planck:2018vyg}, 
which is a clear evidence of Charge-Parity ($C\!P$) violation and Charge-conjugate ($C$) violation, 
according to Sakharov's criteria for the generation of the Baryon Asymmetry of the Universe \cite{Sakharov:1967dj}.
$C\!P$ violation (CPV) has been observed in the decays of $K$ \cite{Christenson:1964fg}, $D$  \cite{LHCb:2019hro}, $B$ \cite{BaBar:2001ags,Belle:2001zzw,LHCb:2013syl}  and $B_s$ \cite{Workman:2022ynf} mesons, all of which are consistent with the description of the Standard Model of particle physics, in which a single $C$-violating weak phase in the Cabibbo–Kobayashi–Maskawa (CKM) matrix is the origin of CPV \cite{Cabibbo:1965zzb,Kobayashi:1973fv}.
Nevertheless, CPV in baryon decay processes is still undiscovered in laboratories to date, despite the fact that many efforts have been made.

In baryon decay processes, the non-zero baryon spin provides us with more freedom to construct 
CPV observables.
The $C\!P$ asymmetries (CPAs) induced by the decay asymmetry parameters in the hyperon weak decay transitions ${\frac{1}{2}}^+\to {\frac{1}{2}}^++0^-$ such as $\Xi \to \Lambda \pi$ and $\Lambda\to p \pi$ are one kind of typical examples.
The decay asymmetry parameters of transitions ${\frac{1}{2}}^+\to {\frac{1}{2}}^++0^-$, are induced by the interference of the parity-even S-wave and the parity-odd P-wave amplitudes \cite{Lee:1957qs}.
They can be measured from the decay-angular-distribution if (1) the mother baryon is polarized, and/or (2) the daughter baryon decays subsequently via weak interactions.
The decay asymmetry parameter induced CPAs in hyperon decays, which have been investigated extensively on the theoretical side \cite{Donoghue:1985ww,Donoghue:1986hh,Bigi:2017eni,Wang:2022tcm}, 
are expected to be larger than those induced by the branching ratios in the aforementioned hyperon decay channels. 
Recently, experimental studies on the hyperon decay asymmetry parameter induced CPAs were performed by BESIII and Belle, with the precision at about one percent level \cite{BESIII:2018cnd,Belle:2022uod}. 
Moreover, decay asymmetry parameter induced CPAs in various charmed baryon decay channels such as $\Lambda_c^+\to (\Lambda, \Sigma^0) h^+$ and $\Sigma_c^0\to \Sigma^- \pi^+$ were also investigated by Belle \cite{Belle:2022uod,Tang:2022ksv}.
However, they are still too small to be confirmed experimentally at the current stage for both the hyperon and the charmed baryon decays.
Theoretical analyses of CPA induced by the decay parameters were also made in bottom baryon decays such as $\Lambda_b^0\to\Lambda D$ \cite{Giri:2001ju,Zhang:2021sit},   
while the corresponding experimental study is still absent currently in bottom baryon decays.

Just like the case of the aforementioned decay asymmetry parameter induced CPAs, the mechanism of generating CPV via the interference between different canonical amplitudes is a good idea for decay processes with particles of non-zero spin involved, and is complementary to CPAs corresponding to the partial decay width.
It should be pointed out, however, that  
the interference between the parity-even and -odd amplitudes  
can only show up in the angular-distributions of the final particles when the subsequent decay is also a weak one. 
In the situation that the corresponding subsequent decay is through strong or electromagnetic interactions which respect parity symmetry, 
the interference between parity-even and -odd amplitudes would be simply absent
\footnote{The interference between parity-even and -odd amplitudes will show up when there are two resonances with similar masses but opposite parities \cite{Zhang:2013oqa,Zhang:2022emj}.}.
Interestingly, for some decay processes, when more than one parity-even and/or -odd amplitudes enter in the decay,  the interference between the amplitudes with the \textit{same parity} can then show up in the decay-angular-distributions, even when the subsequent decay is a strong one. 
We will analyse this in more detail in this paper.

This paper is organized as follows.
In Sec. \ref{sec:gaDADCPA}, we present the general analysis of the decay-angular-distribution correlated CPAs.
In Sec. \ref{sec:DADinCDA}, by expressing the decay-angular-distribution in terms of the canonical decay amplitudes, the interfering behaviour between different canonical decay amplitudes can be seen in a more transparent way.
In Sec. \ref{sec:channels}, we present some suggested decay channels in which the decay-angular-distribution correlated CPAs are suitable for searching for.
In the last section, we make our conclusion.

\section{\label{sec:gaDADCPA} General analysis of the Decay-angular-distributions and the correlated CPAs}
In general, we will consider a mother hadron $\mathbb{H}$ decaying via a weak decay process $\mathbb{H}\to R c$ with the intermediate resonance $R$ decaying through strong interactions $R\to a b$.
The differential decay width of the aforementioned cascade decay process $\mathbb{H}\to R(\to ab) c$
for unpolarized ${\mathbb{H}}$ can be expressed as \cite{Workman:2022ynf}
\footnote{One reason why we only consider the unpolarized case is that the polarizations of the heavy baryons produced on colliders are still too small to be detected \cite{LHCb:2013hzx,LHCb:2020iux,CLEO:1990unw}.}, 
\begin{equation}\label{dGammaMJ}
\frac{d\Gamma_{\mathbb{H}\to R(\to ab) c}}{d s_{ab} ds_{ac}}= \frac{1}{(2\pi)^3}\frac{1}{32m_{\mathbb{H}}^3}\overline{\left|\mathcal{M}\right|^2},
\end{equation}
where $s_{ij}$ is the invariant mass squared of the particle $i$ and particle $j$ system, $m_{\mathbb{H}}$ is the mass of $\mathbb{H}$, and $\overline{\left|\mathcal{M}\right|^2}$ is the decay amplitude squared of the cascade decay $\mathbb{H}\to R(\to ab) c$, which is defined as
$\overline{\left|\mathcal{M}\right|^2}\equiv \frac{1}{2s_{\mathbb{H}}+1}\sum_{m_z,\lambda_a,\lambda_b,\lambda_c}\overline{\left|\mathcal{M}^{s_{\mathbb{H}},m_z}_{\lambda_a\lambda_b\lambda_c}\right|^2}$, 
where $\mathcal{M}^{s_{\mathbb{H}},m_z}_{\lambda_a\lambda_b\lambda_c}$ is the covariant decay amplitude for the cascade decay $\mathbb{H}\to R(\to ab) c $, $s_{\mathbb{H}}$ is the spin of $\mathbb{H}$ and $m_z$ is its $z$-component (the direction of $z$ is irrelevant here since we are dealing with unpolarized $\mathbb{H}$), $\lambda_a$, $\lambda_b$, and $\lambda_c$ are the helicities of $a$, $b$, and $c$, respectively.
The decay amplitude squared can be further expressed as \cite{Zhang:2022emj}
\begin{equation}\label{MJ}
\overline{\left|\mathcal{M}\right|^2}=\sum_{\stackrel{0\leq j\leq 2 {s_{R}}}{j ~\text{even}}} w^{(j)} P_j\left(c_{\theta}\right),
\end{equation}
where $P_j$ is the $j$-th Legendre polynomial, $c_\theta\equiv \cos\theta$ with $\theta$ being the helicity angle of particle $a$ with respect to $c$ (or equivalently, to $\mathbb{H}$) in the center-of-mass frame of the $a \& b$ system, $s_R$ is the spin quantum number for resonance $R$.
Note that $j$ can only take even values (from 0 to $2s_R$), because  the decay of $R$ is through strong interactions which respect the parity symmetry 
 \cite{Zhang:2022emj}.
 
Obviously, all the weights $w^{(j)}$ descibe the decay-angular-distributions, i.e., the angular distributions of the final particles.
The weight $w^{(j)}$ for the $j$-th Legendre polynomial in Eq. (\ref{MJ}) can be expressed as
\begin{eqnarray}\label{wj}
  w^{(j)}
  = \frac{{\langle s_{R}, -s_R; s_{R},s_R|s_{R} s_{R}j0\rangle}^2}{\left|s_{ab}-m_R^2+im_R\Gamma_R\right|^2}\mathcal{W}^{(j)}\mathcal{S}^{(j)},
\end{eqnarray}
where $\mathcal{W}^{(j)}$ and $\mathcal{S}^{(j)}$ contains the information for the decay $\mathbb{H}\to R c$ and $R\to ab$, respectively, and can be expressed in terms of the helicity decay amplitudes as
\begin{eqnarray}\label{eq:Wjhelicity}
  \mathcal{W}^{(j)}&=&\sum_{\sigma}\frac{(-)^{\sigma-s_R}\langle s_{R}, -\sigma; s_{R},\sigma|s_{R} s_{R}j0\rangle}{\langle s_{R}, -s_R; s_{R},s_R|s_{R} s_{R}j0\rangle} \sum_{\lambda_c}\left|\mathcal{F}^{s_\mathbb{H}}_{\sigma\lambda_c}\right|^2,
\end{eqnarray}
and
\begin{eqnarray}\label{eq:Sj}
  \mathcal{S}^{(j)}&=&\left.\sum_{\lambda_a\lambda_b}\frac{(-)^{-\lambda+s_R}\langle s_{R},-\!\lambda; s_{R},\lambda|s_{R}s_{R}j0\rangle} {\langle s_{R}, -s_R; s_{R},s_R|s_{R} s_{R}j0\rangle} \left|\mathcal{G}^{s_{R}}_{\lambda_a\lambda_b}\right|^2\right|_{\lambda=\lambda_a-\lambda_b},
\end{eqnarray}
with $\mathcal{F}^{s_{\mathbb{H}}}_{\sigma\lambda_c}$ and $\mathcal{G}^{s_{R}}_{\lambda_a\lambda_b}$ being the helicity decay amplitudes of ${\mathbb{H}}\to R c$ and $R\to ab$, respectively, $\sigma$ being the helicity of $R$, ``$\langle\cdots|\cdots\rangle$'' being the notation for Clebsch-Gordan coefficients. 
Our definitions of $\mathcal{W}^{(j)}$ and $\mathcal{S}^{(j)}$ are slightly different from those in previous works such as Ref. \cite{Zhang:2022emj}, i.e., with an extra factor $\frac{1}{\langle s_{R}, -s_R; s_{R},s_R|s_{R} s_{R}j0\rangle}$. 
The motivation for the introduction of this factor is to guarantee that $\mathcal{W}^{(0)}=\sum_{\sigma,\lambda_c}\left|\mathcal{F}^{s_\mathbb{H}}_{\sigma\lambda_c}\right|^2$ and $\mathcal{S}^{(0)}=\sum_{\lambda_a\lambda_b} \left|\mathcal{G}^{s_{R}}_{\lambda_a\lambda_b}\right|^2$. 
Moreover, one can see that all the coefficients in front of the helicity amplitudes squared in $\mathcal{W}^{(j)}$ and $\mathcal{S}^{(j)}$ are now simple rational numbers.
One can relax the constraint of $j$ taking even values in Eq. (\ref{MJ}), because $\mathcal{S}^{(j)}$ automatically equals to 0 for odd $j$ due to the parity symmetry in $R\to ab$ \cite{Zhang:2022emj}.

We will focus on the decay-angular-distributions with respect to the helicity angle $\theta$.
To this end, we need to integrate out $s_{ab}$. 
In the narrow width approximation of $R$, the differential decay width of the cascade decay $\mathbb{H}\to R(\to ab) c$ can then be expressed as
\begin{eqnarray}
\frac{1}{\Gamma_{\mathbb{H}\to R(\to ab) c}}\frac{d\Gamma_{\mathbb{H}\to R(\to ab) c}}{dc_{\theta}} =\frac{1}{2} \sum_{\stackrel{0\leq j\leq 2 {s_{R}}}{j ~\text{even}}} \gamma_{\mathbb{H}\to R(\to ab) c}^{(j)}P_j(c_\theta),
\end{eqnarray}
where
$\gamma_{\mathbb{H}\to R(\to ab) c}^{(j)} $ can be further parameterized as $\gamma_{\mathbb{H}\to R(\to ab) c}^{(j)}={\lambda}^{(j)}_{\mathbb{H}\to R c} \alpha^{(j)}_{R\to ab}$, with ${\lambda}^{(j)}_{\mathbb{H}\to R c}$ and $\alpha^{(j)}_{R\to ab}$ 
being defined as
\begin{equation}
  {\lambda}^{(j)}_{\mathbb{H}\to R c}\equiv \frac{{\langle s_{R}, -s_R; s_{R},s_R|s_{R} s_{R}j0\rangle}}{{\langle s_{R}, -s_R; s_{R},s_R|s_{R} s_{R}00\rangle}}\frac{\mathcal{W}^{(j)}}{\mathcal{W}^{(0)}},
\end{equation}
and 
\begin{equation}
  \alpha^{(j)}_{R\to ab}= \frac{{\langle s_{R}, -s_R; s_{R},s_R|s_{R} s_{R}j0\rangle}}{{\langle s_{R}, -s_R; s_{R},s_R|s_{R} s_{R}00\rangle}} \frac{\mathcal{S}^{(j)}}{\mathcal{S}^{(0)}},
\end{equation}
respectively.
Here, $\alpha^{(j)}_{R\to ab}$ are a set of decay parameters of $R\to ab$.
On the other hand, ${\lambda}^{(j)}_{\mathbb{H}\to R c}$ ($j=1,\cdots,2s_R$) describe the polarization of $R$, as can be seen from Eq. (\ref{eq:Wjhelicity})
\footnote{
We can further parameterize ${\lambda}^{(j)}_{\mathbb{H}\to R c}$ as
${\lambda}^{(j)}_{\mathbb{H}\to R c}=\mathcal{P}^{(j)}_{R}\kappa^{(j)}_{\mathbb{H}\to R c}$.
where $\mathcal{P}^{(j)}_{R}$ are the normalized polarization parameters which are defined as
$\mathcal{P}^{(j)}_{R}\equiv\frac{\mathcal{W}^{(j)}}{\mathcal{W}^{(j)}_{\text{abs}}}$,
and $\kappa^{(j)}_{\mathbb{H}\to R c}$ can be viewed as production asymmetry/ratio and are defined through
$\kappa^{(j)}_{\mathbb{H}\to R c}\equiv\frac{\mathcal{W}^{(j)}_{\text{abs}}}{\mathcal{W}^{(0)}}$,
where
$\mathcal{W}^{(j)}_{\text{abs}}\equiv \sum_{\sigma,\lambda_c}\left| \frac{(-)^{\sigma-s_R}\langle s_{R}, -\sigma; s_{R},\sigma|s_{R} s_{R}j0\rangle}{\langle s_{R}, -s_R; s_{R},s_R|s_{R} s_{R}j0\rangle}\right| \left|\mathcal{F}^{s_\mathbb{H}}_{\sigma\lambda_c}\right|^2$.
}.
They can be viewed as the generalization of the polarization from spin-half particles to particles of any spins produced in weak decay processes.
We say that $R$ is unpolarized if ${\lambda}^{(j)}_{\mathbb{H}\to R c}=0$, for all $j=1,2,\cdots,2s_R$.

The differences between $\lambda_{\mathbb{H}\to R c}^{(j)}$ and their $CP$ correspondence $\lambda_{\overline{\mathbb{H}}\to \overline{R} \overline{c}}^{(j)}$, are measures of $CP$V. 
One can define the decay-angular-distribution correlated CPAs as \cite{Zhang:2021fdd,Wei:2022zuf,Zhang:2022iye}
\begin{equation}\label{eq:ACPj}
  {A}_{CP}^{(j)}\equiv
  \frac{1}{2}\left(\lambda_{\mathbb{H}\to R c}^{(j)}-\lambda_{\overline{\mathbb{H}}\to \overline{R} \overline{c}}^{(j)}\right).
\end{equation}
Note that  ${A}_{CP}^{(0)}\equiv0$ by definition. 
From the experimental side, the direct observables are
\begin{equation}
  \tilde{A}_{CP}^{(j)}={A}_{CP}^{(j)}\alpha^{(j)}_{R\to ab},
\end{equation}
from which one can see that the observability of ${A}_{CP}^{(j)}$ are restricted by the values of $\alpha^{(j)}_{R\to ab}$.
In other words, $\tilde{A}_{CP}^{(j)}$ can be observed only when both ${A}_{CP}^{(j)}$ and $\alpha^{(j)}_{R\to ab}$ are not too small.

In practice, the following definition of CPA observables is easier accessible experimentally. 
The null points of $P_j(c_\theta)$ devide the interval of $c_\theta\in [-1,1]$ into $j+1$ sub-intervals, which can be ordered from right ($c_\theta=+1$) to left ($c_\theta=-1$) as the 1st, 2nd, $\cdots$, $k$-th, $\cdots$, $(j+1)$-th sub-intervals. Denoting the invent yields of each sub-interval as $N^{(j)}_k$ and $\overline{N^{(j)}_k}$ for $k=1,\cdots,j+1$, the CPA observables can be defined as
\begin{equation}\label{eq:ACPj}
  \hat{A}_{CP}^{(j)}\equiv\frac{\sum_{k=1}^{j+1} (-)^k\left[ N_k^{(j)}- \overline{N_k^{(j)}}\right]}{\sum_k \left[ N_k^{(j)}+ \overline{N_k^{(j)}}\right]}.
\end{equation}
It can be easily seen that $\hat{A}_{CP}^{(0)}$ is in fact the conventionally defined CPA corresponding to the branching ratios, ${A}_{CP} (\mathbb{H}\to R (\to ab)c)$.

\section{\label{sec:DADinCDA} Interference behaviour of canonical decay amplitudes in the Decay-angular-distributions} 

 Interference between amplitudes with different weak phases may generate large a $CP$ asymmetry parameter, provided that there is also a large strong phase difference.
Potential final state interaction may generate such non-perturbative strong phase difference in different canonical partial wave amplitudes \cite{Watson:1952ji}. 
Besides, the interference behaviour is quite obscure in Eq. (\ref{eq:Wjhelicity}) since there are no interfering terms between different helicity amplitudes for the current situation.
Consequently, it will be more transparent to see the interference pattern if we re-express $\mathcal{W}^{(j)}$ in terms of canonical amplitudes,
which reads
  \begin{eqnarray}
  \mathcal{W}^{(j)}
&=&\sum_{ls,l's'} \rho_{ls,l's'}^{(j)} a^{s_{\mathbb{H}}}_{ls} a^{s_{\mathbb{H}}\ast}_{l's'},
  \end{eqnarray}
where $a^{s_{\mathbb{H}}}_{ls}$'s are the canonical decay amplitudes of ${\mathbb{H}}\to R c$ and are related to the helicity amplitudes through \cite{Jacob:1959at,Chung:1971ri}
\begin{equation}
\mathcal{F}^{s_{\mathbb{H}}}_{\sigma\lambda_3}
  =\sum_{ls} \left(\frac{2l+1}{2 s_{\mathbb{H}}+1}\right)^{\frac{1}{2}} \langle s_{R},\sigma; s_c, -\!\lambda_c|s_{R} s_c s (\sigma\!\!-\!\!\lambda_c)\rangle \langle l, 0; s, \sigma\!\!-\!\!\lambda_c|ls s_{\mathbb{H}}(\sigma\!\!-\!\!\lambda_c)\rangle  a^{s_{\mathbb{H}}}_{ls},
\end{equation}
or inversely,
\begin{equation}
a^{s_{\mathbb{H}}}_{ls}
  =\sum_{\sigma\lambda_c} \left(\frac{2l+1}{2 s_{\mathbb{H}}+1}\right)^{\frac{1}{2}} \langle s_{R},\sigma; s_c, -\!\lambda_c|s_{R} s_c s (\sigma\!\!-\!\!\lambda_c)\rangle \langle l, 0; S, \sigma\!\!-\!\!\lambda_c|ls s_{\mathbb{H}}(\sigma\!\!-\!\!\lambda_c)\rangle \mathcal{F}^{s_{\mathbb{H}}}_{\sigma\lambda_c},
\end{equation}
and the rotational-invariant coefficient $\rho_{ls,l's'}^{(j)}$ reads
\begin{eqnarray}
  \rho_{ls,l's'}^{(j)}&=& \frac{\sqrt{(2l+1)(2l'+1)}}{(2 s_{\mathbb{H}}+1)\langle s_{R}, -s_R; s_{R},s_R|s_{R} s_{R}j0\rangle}
  \sum_{\sigma\lambda_c}(-)^{\sigma-s_R} \langle s_{R},-\!\sigma; s_{R},\sigma|s_{R}s_{R}j0\rangle \nonumber\\
  &\times& \langle l,0; s, {\sigma\!\!-\!\!\lambda_c}|l s s_{\mathbb{H}} (\sigma\!\!-\!\!\lambda_c)\rangle 
  \langle s_{R},\sigma; s_c,-\!\lambda_c|s_{R} s_c s (\sigma\!\!-\!\!\lambda_c)\rangle 
  \nonumber\\
  &\times& 
  \langle l',0;
  s',(\sigma\!\!-\!\!\lambda_c)|l's' s_{\mathbb{H}} (\sigma\!\!-\!\!\lambda_c)\rangle \langle s_{R},\sigma; s_c,-\!\lambda_c|s_{R}s_c s' (\sigma\!\!-\!\!\lambda_c)\rangle.
  \end{eqnarray}

In general, the presence of the interference between two canonical decay amplitudes in certain $\mathcal{W}^{(j)}$ is constrained by the properties of the coefficients $\rho_{ls,l's'}^{(j)}$.
The following two properties of the coefficients $\rho_{ls,l's'}^{(j)}$ , which can be proven with the aid of the properties of the Clabsch-Gordan coefficients, 
turn out to be very important for our analysis here: 
\begin{enumerate}

\item {For a given value of $j$ satisfying $0\leq j \leq 2s_R$, the indices of the nonzero elements of $\rho_{ls,l's'}^{(j)}$ fulfill the triangle inequality (necessary condition): }
\begin{equation} \label{eq:lj}
|l-l'|\leq j\leq l+l',
\end{equation}  
\begin{equation}\label{eq:sj}
|s-s'|\leq j\leq s+s'.
\end{equation}
\item { Zero elements:} 
 \begin{equation}\label{eq:lsj}
  \rho_{ls,l's'}^{j}=0, ~\text{if} \left\{ 
  \begin{array}{l}
                                                                    \text{$j$ is even, one of $l$ and $l'$ is even and the other is odd;} \\
                                                                    \text{$j$ is odd, both $l$ and $l'$ are even or odd.}
                                                                  \end{array}
                                                                  \right.
  \end{equation} 
\end{enumerate}

The first consequence of Property 2 is that ---since $j$ can only take even values in Eq. (\ref{MJ})---  the interference between parity-even and -odd amplitudes
 is absent in the decay amplitude squared of Eq. (\ref{MJ}) when the subsequential decay of $R$ is strong, where the parity of the canonical decay amplitude $a^{s_\mathbb{H}}_{ls}$ is determined according to $\Pi_{a^{s_\mathbb{H}}_{ls}}=(-)^l \Pi_{\mathbb{H}}\Pi_{R}\Pi_{c}$, with $\Pi$ representing parity. 
This means that there can be only interference of $a^{s_{\mathbb{H}}}_{ls}$ and $a^{s_{\mathbb{H}}}_{l's'}$ in $\mathcal{W}^{(j)}$ of Eq. (\ref{MJ}) when both $l$ and $l'$ are even or odd simultaneously for even $j$ (so that $a^{s_{\mathbb{H}}}_{ls}$ and $a^{s_{\mathbb{H}}}_{l's'}$ have the same parity).

Of course, Property 1 can also set constraints on the presence of the interfering terms in $\mathcal{W}^{(j)}$.
For example, there will be no interfering terms between different canonical amplitudes in $\mathcal{W}^{(0)}$, because according to Property 1, $l=l'$ and $s=s'$ for $j=0$.
Hence one has $\rho_{ls,l's'}^{0}=\delta_{ll'}\delta_{ss'}$ when $j=0$, so that $\mathcal{W}^{(0)}=\sum_{\sigma\lambda_c}\left|\mathcal{F}_{\sigma\lambda_c}^{s_{\mathbb{H}}}\right|^2 =\sum_{ls}\left|a_{ls}^{s_\mathbb{H}}\right|^2$, from which one can see that there is no interference between different canonical amplitudes in $w^{(0)}$, as expected.

For the weak process $\mathbb{H}\to R c$, it can be proved that the number of the independent canonical decay amplitudes, which is of course the same as that of the independent helicity decay amplitudes, are in total 
\begin{equation}\label{eq:Nca}
N_{c.a.}=(2 s_{1}+1)(2 s_2+1)-\kappa(\kappa+1), 
\end{equation}
where $\kappa\equiv\max\left\{s_1+s_2-s_3,0\right\}$.
Here, $s_1$, $s_2$, and $s_3$ represent the spins of the particles involved in the weak decay, $\mathbb{H}$, $R$, and $c$, which are ordered according to $s_1\leq s_2\leq s_3$.
Moreover, it can be proved that the number of parity-even and -odd canonical amplitudes will be either the same (when $N_{c.a.}$ is a even number), or with a difference of 1 (when $N_{c.a.}$ is an odd number).
Consequently, in order for the interferences between the canonical decay amplitudes with the same parity in $w^{(j)}$ Eq. (\ref{MJ}) for certain $j$ to happen, the number of the independent canonical decay amplitudes should be no less than 3, so that at least two of them share the same parity. 
The first few combinations of $(s_1,s_2,s_3)$ that fulfill this requirement are $(0,1,1)$, $(0,1,2)$, $(0,\frac{3}{2},\frac{3}{2})$, $(0,2,2)$, $(\frac{1}{2},\frac{1}{2},1)$, $(\frac{1}{2},\frac{1}{2},2)$, $(\frac{1}{2}, 1, \frac{3}{2})$, and $(\frac{1}{2}, \frac{3}{2}, 2)$, with the number of independent canonical decay amplitudes of 3, 3, 4, 5, 4, 4, 6, and 8, respectively.
Phenomenologically, if we confine ourself only on discussions of unstable $\mathbb{H}$ that is dominated by weak decays, the first four combinations will be the cases of pseudo-scalar heavy meson decays, while the last four combinations will be the cases of heavy baryon decays. 

Take the typical weak transition of the type $\mathbb{B}\to \mathcal{B} V$ (${\frac{1}{2}}^+\to {\frac{1}{2}}^+ + 1^-$), such as $\Lambda_b\to\Lambda \rho^0$,
 as an example, where the vector meson decays subsequently to two pseudo-scalar mesons through strong interactions.
It can be easily seen that there are four independent weak decay canonical amplitudes in total, according to Eq. (\ref{eq:Nca}). 
There are in total three $\mathcal{W}^{(j)}$'s for $j=0,1$, and $2$, which can be respectively expressed in terms of the helicity amplitudes as well as the canonical ones as 
\begin{eqnarray}
\label{eq:w0PWh}
\mathcal{W}^{(0)} &=&|\mathcal{F}_{1 \frac{1}{2}}|^2+|\mathcal{F}_{\!-1 -\frac{1}{2}}|^2+|\mathcal{F}_{0 \frac{1}{2}}|^2+|\mathcal{F}_{0 ~-\frac{1}{2}}|^2,\\
\label{eq:w0PWh}
\mathcal{W}^{(1)} &=&|\mathcal{F}_{1 \frac{1}{2}}|^2-|\mathcal{F}_{\!-1 -\frac{1}{2}}|^2,\\
\label{eq:w2PWh}
\mathcal{W}^{(2)}&=&|\mathcal{F}_{1 \frac{1}{2}}|^2+|\mathcal{F}_{\!-1 -\frac{1}{2}}|^2-2\left(|\mathcal{F}_{0 \frac{1}{2}}|^2+|\mathcal{F}_{0 ~-\frac{1}{2}}|^2\right), 
\end{eqnarray}
and
\begin{eqnarray}
\label{eq:w0PWc}
\mathcal{W}^{(0)} &=&\left(|a_{0\frac{1}{2}}|^2+|a_{1\frac{1}{2}}|^2+|a_{1\frac{3}{2}}|^2+|a_{2\frac{3}{2}}|^2\right),\\
\label{eq:w0PWc}
\mathcal{W}^{(1)}
&=&\frac{2}{\sqrt{3}}\left[\sqrt{2}\Re\left(a_{0\frac{1}{2}}a_{1\frac{1}{2}}^\ast  \right) +\Re\left(a_{0\frac{1}{2}}a_{1\frac{3}{2}}^\ast  \right) +\Re\left(a_{1\frac{1}{2}}a_{2\frac{3}{2}}^\ast  \right)+\sqrt{\frac{1}{2}}\Re\left(a_{1\frac{3}{2}}a_{2\frac{3}{2}}^\ast  \right)\right],\\
\label{eq:w2PWc}
\mathcal{W}^{(2)}
&=& \frac{1}{\sqrt{2}}\left[ -\left(|a_{1\frac{3}{2}}|^2+|a_{2\frac{3}{2}}|^2\right)+2\sqrt{2}\Re\left(a_{0\frac{1}{2}}a_{2\frac{3}{2}}^\ast  +a_{1\frac{1}{2}}a_{1\frac{3}{2}}^\ast\right)\right].
\end{eqnarray}
From the helicity forms of $\mathcal{W}^{(1)}$ and $\mathcal{W}^{(2)}$ one can clearly see that they describe the polarization of the vector meson $V$:
$\mathcal{W}^{(1)}$ represents the asymmetry between helicity $+1$ and $-1$ of $V$; while $\mathcal{W}^{(2)}$ represents the asymmetry of transverse and longitudinal polarizations. 
The factor 2 in front of the longitudinal polarization parts is important, and also understandable. 
It reflects the simple fact that the degrees of freedom for the longitudinal polarizations ($\sigma=\pm1$) and the transverse parts ($\sigma=0$) are different. 
On the other hand, the interference behaviour between different amplitudes is easier to be seen from the canonical forms of $\mathcal{W}^{(1)}$ and $\mathcal{W}^{(2)}$.

The differential decay width can then be expressed as
\footnote{The behaviour in the expression seems to be different from the one used in the literature such as \cite{Pakvasa:1990if}. One can see from Eq. (13) of Ref. \cite{Pakvasa:1990if} that there are interference terms of S- and D-waves in the integrated decay width $\Gamma$, while there is no such kind of term (interference between $a_{0\frac{1}{2}}$ and $a_{2\frac{3}{2}}$ in our notation) in Eq. (\ref{eq:w0PWc}). This is because the definitions of the canonical amplitudes are different. In the current paper, $a_{ls}$ are the canonical amplitudes in the sense that they transform irreducibly under SO(3), while the $S$- and $D$- waves in Ref. \cite{Pakvasa:1990if} are not.}
\begin{equation}
\frac{1}{\Gamma}\frac{d\Gamma}{d\cos\theta}=\frac{1}{2}+\frac{\mathcal{W}^{(2)}{\mathcal{S}}^{(2)}}{2\mathcal{W}^{(0)}{\mathcal{S}}^{(0)}}P_2(c_\theta)
=\frac{1}{2}-\frac{1}{\sqrt{2}}\lambda^{(2)}_{\mathbb{B}\to \mathcal{B} V}P_2(c_\theta).
\end{equation}
Note that $\mathcal{S}^{(0)}$ and $\mathcal{S}^{(2)}$ are correlated according to $\mathcal{S}^{(2)}/\mathcal{S}^{(0)}=-\sqrt{2}$.
The reason for the correlation is understandable. Since the vector meson decays into two pesudoscalar ones, there can only be one independent helicity amplitude. Both of $\mathcal{S}^{(0)}$ and $\mathcal{S}^{(2)}$ must proportional to the square of this helicity amplitude, hence they are correlated.
If the relative strong phase differences between $a_{0\frac{1}{2}}$ and $a_{2\frac{3}{2}}$ and/or $a_{1\frac{1}{2}}$ and $a_{1\frac{3}{2}}$ are not small, the contributions of the corresponding interfering terms to the CPA associated with the decay-angular-distributions $A_{CP}^{(2)}$ could be large.

To see the CPA behaviour in more detail, let us first parameterize the canonical decay amplitudes $a_{ls}^{s_{\mathbb{H}}}$ and their $C\!P$ counterparts $\overline{a_{ls}^{s_{\mathbb{H}}}}$ as 
\begin{equation}
  a_{ls}^{s_{\mathbb{H}}}=\left(a_{ls}^T +a_{ls}^P e^{i(\phi_{\text{w}}+\delta_{ls}^{PT})}\right)e^{i\delta_{ls}},
\end{equation}
\begin{equation}
  \overline{a_{ls}^{s_{\mathbb{H}}}}=\left(a_{ls}^T +a_{ls}^P e^{i(-\phi_{\text{w}}+\delta_{ls}^{PT})}\right)e^{i\delta_{ls}},
\end{equation}
where $a_{ls}^T$ and $a_{ls}^P$ are the tree and penguin parts of the canonical decay amplitudes, respectively, $\phi_{\text{w}}$ and $\delta_{ls}^{PT}$ are respectively the weak and the strong phase difference between the tree and penguin parts of the same canonical decay amplitude $a_{ls}$,  and $\delta_{ls}$ is the overall strong phase of the canonical amplitudes $a_{ls}^{s_{\mathbb{H}}}$.
The CPA parameters will behave as 
\begin{equation}
  A^{(j)}_{CP}\sim 2\sin\phi_{\text{w}}\sum_{ls,l's'}\rho^{(j)}_{ls,l's'}\left\{a^T_{ls}a^P_{l's'}\sin\left[\left(\delta_{ls}-\delta_{l's'}\right)-\delta_{l's'}^{PT}\right] -a^P_{ls}a^T_{l's'}\sin\left[\left(\delta_{ls}-\delta_{l's'}\right)+\delta_{ls}^{PT}\right]\right\}.
\end{equation}
It is naturally expected (although this is only a possibility) that the relative strong phases between different canonical decay amplitudes could be relatively larger, comparing to the strong phase difference of the tree and the penguin parts with different weak phases within the same canonical decay amplitude,
\begin{equation}
  \delta_{ls}-\delta_{l's'}>\delta_{ls}^{PT}\sim\delta_{l's'}^{PT}.
\end{equation}
This means that the CPA parameter(s) $A^{(j)}_{CP}$ will have a good chance to be dominated by the interference terms between different canonical decay amplitudes:
\begin{equation}
  A^{(j)}_{CP}\sim 2\sin\phi_{\text{w}}\sum_{ls\neq l's'}\rho^{(j)}_{ls,l's'}\left(a^T_{ls}a^P_{l's'}-a^P_{ls}a^T_{l's'}\right)\sin\left(\delta_{ls}-\delta_{l's'}\right).
\end{equation}
Since there is no interference between different canonical decay amplitudes in $\hat{A}^{(0)}_{CP}$, it is probably that $A^{(j)}_{CP}$ for $j\neq 0$ will be larger than $\hat{A}^{(0)}_{CP}$.

While the presence of the strong phase difference between different partial-wave amplitudes is an important necessary condition for large CPA corresponding to the decay-angular distributions, it should be pointed out that this is not a sufficient condition. 
To see this, let us use again the decay $\mathbb{B}\to \mathcal{B} V$ as an example to illustrate.
Theoretical analyses of this type of decays have been performed via Generalized Factorization Approach (GFA) for bottom baryon decays \cite{Leitner:2006sc,Geng:2020zgr,Geng:2021sxe}.
Based on helicity decay amplitudes obtained from GFA, 
it can be shown that the canonical decay amplitudes, and their $C\!P$ conjugates, take the forms 
$a_{ls}=\mathcal{C}_V K_{ls}$, and $\overline{a_{ls}}=\overline{\mathcal{C}_V} K_{ls}$, where $K_{ls}$ is the kinematical part which depends on $l$ and $s$, and  the parameters $\mathcal{C}_V$ and $\overline{\mathcal{C}_V}$ contain the CKM matrix elements and are universal for all the $a_{ls}$.
Especially, $\mathcal{C}_V$ and $\overline{\mathcal{C}_V}$ are independent of $l$ and $s$.
While it can generate a non-zero direct CPA corresponding to the branching ratios, the CPAs corresponding to the decay-angular-distributions, which are defined in Eq. (\ref{eq:ACPj}), are predicted to be exactly zero by GFA, because the universal $\mathcal{C}_V$ ($\overline{\mathcal{C}_V}$) is cancelled out in $\lambda_{\mathbb{H}\to R c}^{(j)}$ ($\lambda_{\overline{\mathbb{H}}\to \overline{R} \overline{c}}^{(j)}$), so that $\lambda_{\mathbb{H}\to R c}^{(j)}=\lambda_{\overline{\mathbb{H}}\to \overline{R} \overline{c}}^{(j)}$. 
In other words, there can be only one CPA in the approach of GFA, i.e., CPA corresponding to the branching ratios, $\tilde{A}_{CP}^{(0)}$.
Consequently, it is crucially important to go beyond GFA for the predictions of CPA corresponding to the decay-angular distributions.

\section{\label{sec:channels}suggested channels for searching for decay-angular-distributions correlated CPAs in baryon decays}

It is very hard to make a concrete prediction of any decay-distribution-correlated CPAs in heavy baryon cascade decays.
Nevertheless, our analysis do provide some guidelines for searching for such kind of CPAs in certain decay processes.
According to the above analysis, two of the guidelines are especially important.
The first one is that the number of independent canonical decay amplitudes $N_{c.a.}$ should be no less than $3$, 
which will constrain the spins of the particle involved according to Eq. (\ref{eq:Nca}). 
The second one is the that $j$ can only take even values from $0$ to $2s_R$, which implies that $s_R$ should be no less than 1 in order for the non-trivial decay-angular-distributions ($j\geq2$) to appear.
Of course, the properties of the coefficients $\rho^{(j)}_{ls,l's'}$ can provide us more detailed information on the interference of different canonical decay amplitudes in $\mathcal{W}^{(j)}$. 

The above guidelines suggest that we can study the decay-angular-distributions and the corresponding $CP$As in the following two immediate situations.
The first one is that $\mathbb{H}$ is a pseudo-scalar heavy meson, which will be denoted as $\mathbb{M}$ for this situation.
From Eq. (\ref{eq:Nca}) one can see in this situation that the necessary condition for $N_{c.a.}\geq3$ is that the spins of both $R$ and $c$ should be non-zero and larger than half.
As aforementioned, typical examples for the spin combinations $(s_1, s_2, s_3)$ of $\mathbb{M}$, $R$ and $c$ that fulfill this requirement are (0, 1, 1), (0, 1, 2), $(0,\frac{3}{2},\frac{3}{2})$, (0, 2, 2).
The simplest decays which can be used to perform the search of decay-angular-distribution correlated CPAs are 1) $\mathbb{M}\to V_1V_2$, with $V_1$ or $V_2$ decaying strongly to two pseudoscalar mesons; 2) $\mathbb{M}\to B_1^\ast B_2^\ast$, with $B_1^\ast$ and $B_2^\ast$ being spin-one-and-a-half baryons and one of them decaying via strong interactions.  
Since our main concern is heavy baryon decay processes, we will not go through the heavy meson decay processes any further \cite{Dunietz:1990cj}.

The second situation, which is our main concern in this paper, is that $\mathbb{H}$ represents a spin-half heavy baryon, which will be denoted as $\mathbb{B}$ in what follows. 
In this case, since either $R$ or $c$ is a baryon, in order for $N_{c.a.}\geq3$, Eq. (\ref{eq:Nca}) indicates that the other particle must be a spin-non-zero meson.
In view of the above constraints, we propose to search for decay-distribution-correlated CPV in cascade decays of the following types: 
1)  ${\mathbb{B}}\to {\mathcal{B}} M$, $M\to M_1M_2$, with the spin of the meson $M$ being nonzero;
2)  ${\mathbb{B}}\to {\mathcal{B}} M$, ${\mathcal{B}}\to {\mathcal{B}}'M'$, with the spin of the baryon resonance $\mathcal{B}$ being larger than $\frac{1}{2}$, and the spin of $M$ being nonzero.
Here, $\mathcal{B}$ and $\mathcal{B}'$ represent light baryons, and $M$, $M'$, $M_1$ and $M_2$ represent light mesons.

The first type seems to be more common and more applicable, among which the most relevant decay type is of the form ${\mathbb{B}}\to {\mathcal{B}} V$, with a subsequent strong decay of the vector resonance $V\to P_1P_2$.
Typical decays of the form ${\mathbb{B}}\to {\mathcal{B}} V$ include 
1) $b\to d u\overline{u}$ transitions: $\Lambda_b^0\to p \rho(770)^+$, $\Lambda_b^0\to N(1520)^\ast \rho(770)^+$; 
2) $b\to s u\overline{u}$ transitions: $\Lambda_b^0\to \Lambda \rho(770)^0$, $\Lambda_b^0\to p {K^\ast}(892)^-$, $\Lambda_b^0\to N(1520) K^\ast$;
3) $c\to ud\overline{d}$ transitions: $\Lambda_c^+\to p \rho(770)^0$, $\Xi_c^+\to p \overline{K^\ast}(892)^0$; 
4) $c\to us\overline{s}$ transitions:
 $\Lambda_c^+\to p \phi$,  $\Lambda_c^+\to\Sigma^+ {K^\ast}(892)^0$. 
 
 For the second type, ${\mathbb{B}}\to \mathcal{B} M$, $\mathcal{B}\to \mathcal{B}'M'$,  where the spin of the intermediate baryon resonance is no less than $\frac{3}{2}$, and the spin of $M$ is no less than 1, typical decays include 1) $c\to ud\overline{d}$ transitions: $\Lambda_c^+\to N(1520)^\ast \rho(770)^0$, $\Xi_c^+\to N(1520)^\ast \overline{K^\ast}(892)^0$; 
2) $c\to us\overline{s}$ transitions:
 $\Lambda_c^+\to N(1520)^\ast \phi$,  $\Lambda_c^+\to\Sigma^+ {K^\ast}(892)^0$; 
3) $b\to d u\overline{u}$ transition:  $\Lambda_b^0\to N(1520)^\ast \rho(770)^+$; 
4) $b\to s u\overline{u}$ transition:  $\Lambda_b^0\to N(1520) K^\ast$.
For most of the cases, the study of $A^{(j)}$ and/or $\hat{A}^{(j)}$ for $j=2$ will be enough since all the spins of the resonances mentioned in the above examples are less than 2.

\section{Summary and Conclusion}

$CP$V in baryon decay processes has not been observed yet.
In this paper, the decay-angular-distribution correlated CPAs for cascade decays of unpolarized heavy hadrons are analyzed.
By expressing the differential decay width in terms of the canonical decay amplitudes, we analyse the general condition for the presence of the interfering terms between different canonical decay amplitudes.
The presence of the interfering terms are important for the generation of large decay-angular-distribution correlated CPAs.

We focus mainly on one typical type of decays, in which the heavy baryon decays weakly into two daughter hadrons, 
with one of them decaying strongly into two granddaughter hadrons.
The analysis indicates that when the two daughter hadrons are both spin-nonzero, there will be at least two parity-even and -odd canonical amplitudes. 
The interference between the canonical amplitudes with the same parity properties will be present in the angular distributions of the final particles.
With a possible large strong phase between different canonical amplitudes, a large CPA corresponding to the decay-angular-distributions may be generated.
We also present some typical decay channels in which the search for such kind of CPAs can be performed.

\begin{acknowledgments}
We thank Prof. Hai-Yang Cheng, Prof. Hsiang-nan Li, and Dr. Chia-Wei Liu for valuable discussions.
This work was supported by National Natural Science Foundation of China under Grants Nos. 12192261 and 12275024, Natural Science Foundation of
Hunan Province under Grants No. 2022JJ30483, and Scientific Research Fund of Hunan Provincial Education Department under Grants No. 22A0319.
\end{acknowledgments}

\bibliography{zzhbib}

\end{document}